\documentclass[10pt, conference, a4paper]{IEEEtran}

\usepackage{cite}

\ifCLASSINFOpdf
  \usepackage[pdftex]{graphicx}
\else
  \usepackage[dvips]{graphicx}
\fi
\usepackage[tight,footnotesize]{subfigure}
\usepackage{url}

\usepackage[colorinlistoftodos, obeyDraft]{todonotes}

\usepackage{amsmath}

\begin{document}
%
\title{Parameterization of the SWIM Mobility Model Using Contact Traces}

\author{\IEEEauthorblockN{Zeynep Vatandas, Manikandan Venkateswaran, Koojana Kuladinithi, Andreas Timm-Giel\\
Institute of Communication Networks, 
Hamburg University of Technology,
Germany\\
Email: \{zeynep.vatandas $|$ manikandan.venkateswaran $|$ koojana.kuladinithi $|$ andreas.timm-giel\}@tuhh.de}

}


%


\maketitle

\begin{abstract}

Opportunistic networks (OppNets) are focused to exploit direct, localised communications which occur in a peer-to-peer manner mostly based on people's movements and their contact durations. Therefore the use of realistic mobility models is critical to evaluate the data dissemination in OppNets. One of the mobility models that is available in OMNeT++ which can be used to mimic human movement patterns is Small Worlds in Motion (SWIM). The SWIM model is based on the intuition that humans often visit nearby locations and if the visited location is far away, then it is probably due to the popularity of the location.

As an alternative to mobility of a node, pairwise contact probabilities are also used to evaluate the data dissemination in OppNets. Pairwise contact probabilities can be used to predict that a node will be met by a particular node. These probabilities can be derived in many ways. One of the ways is to calculate the average probability with which a node will meet another particular node at any point of time. Another way is to calculate the probability with which a node will meet another based on the time of day. The way of calculating pairwise contact probability depends on the scenario. 

In this work,  the pairwise contact probabilities obtained from the real traces are used to tune the parameters of the SWIM mobility model. The traces and the SWIM model are compared in terms of contact durations, inter-contact times and, number of pairwise contacts. How to decide SWIM parameters using real contact traces are being addressed.  
   

\end{abstract}


%
\IEEEpeerreviewmaketitle

\section{Introduction}
\label{sec:intro}

Opportunistic networks use the mobility of users (or nodes) to send data by storing and forwarding techniques.  The data transfer is done when there is a contact between two users. In other words, data is transmitted only when two nodes are within their communication range. The communications are often between short range nodes and, ad-hoc in nature. Since the network is delay tolerant, the data transfers are often used for non-critical situations.  When evaluating the performance of OppNets, OMNeT++ plays a major role as a network simulator due to the scalability and availability of different OppNets routing and mobility models. \cite{OPS:2017}

This paper mainly deals with the simulation of opportunistic networks using human mobility models. To simulate human mobility, lots of different mobility models are proposed till now.  Since pure random models are not good and realistic, traces are difficult to obtain, the authors of \cite{Mei:2009} have proposed a simple self tuning mathematical model to model human behaviour, called as Small Worlds in Motion (SWIM) which is based on location preferences alone. In this paper, data from real life traces taken by students from the Cambridge University \cite{Cambridge:2005} are analysed and inferences from the traces are used to decide on the parameters for simulating the SWIM model. The Cambridge traces are Bluetooth traces recorded at various scenarios and locations. Moreover, methods to calculate pairwise contact probabilities are analysed and the influence of the pairwise contact probabilities on choosing the simulation parameters are discussed. The pairwise contact probabilities are represented as a matrix and ways to calculate the matrix are discussed. Based on the parameters derived from traces, the SWIM model is used in our scenarios to compare different performance metrics (such as Inter-contact times, contact durations, number of pairwise contacts, number of contacts per unit time and, pairwise contact probabilities). By comparing these features related to mobility patterns of humans, properties present (and not present) in the SWIM model are discussed. Therefore, the work presented in this paper provides how SWIM parameters can be adjusted to mimic the same characterestics of contact based traces in a given scenario. We use the SWIM mobility model \cite{SWIM:2016} available in OMNeT++ to validate SWIM parameters referring three different traces. 

The rest of this paper is ordered in the following manner. The next section (Section~\ref{sec:swim_traces}) provides a brief introduction to the SWIM mobility model and the contact traces. Section~\ref{sec:scenario} provides how to select SWIM parameters and description of the scenario. Section~\ref{sec:swim_eval} is a validation of the results taken using the SWIM mobility model and the pure contact traces. Section~\ref{sec:conclusion} is a concluding summary.

\section{Overview: SWIM Mobility Model and Contact Traces}
\label{sec:swim_traces}

\subsection{SWIM Mobility Model}

SWIM \cite{Mei:2009} is a simple mobility model meant for efficient simulation of human movements. The model is based on the following two intuitions of human movements. 

\begin{itemize}
	\item A visited location is either near to a person's home location;
	\item or, if the visited location is far from the home location, it is visited due to the popularity of that location.
\end{itemize}

Each node in SWIM has a home location permanently assigned to itself. The home location is chosen randomly from the simulation network area. The nodes can move only to certain number of locations which are scattered around the network area randomly. Other than these locations, a node can move to its own home and each location is treated as a square cell C. Each node maintains a weight w(C) value for locations in the map. A node does not need to have a weight for each and every location in the map. A node only maintains weights for the locations it has visited. At the beginning, the weights will be initialised to 0. The weight of a location can be calculated as shown in equation (\ref{eq:weight}) below.

\begin{equation}\label{eq:weight}
w(C) = \alpha \cdot distance(h_{A}, C) + (1 - \alpha) \cdot seen(C)
\end{equation}

The equation represents the weight that node $A$ assigns to cell $C$. In this equation, $distance(h_{A}, C)$  is a measure  which decays based on power-law of  distance from node $A$'s home to cell $C$, $\alpha$ is a constant value in the range of between 0 and 1 and $seen(C)$ is the number of  encountered nodes at cell $C$ by node $A$ and $seen(C)$ is updated each time node $A$ visits cell $C$. The value of $\alpha$ influences the next destination chosen. If the value of $\alpha$ is large, then a mobile node is more likely to choose a destination near to it's home location while a small $\alpha$ results in the node selecting popular locations away from the home location.

According to \cite{SWIM:2016}, SWIM implementation in OMNeT++ is done by extending the \emph{LineSegmentsMobilityBase} class.

\subsection{Contact Traces}

The SWIM model was simulated by the authors in \cite{Mei:2009} and compared with Cambridge Bluetooth traces \cite{Cambridge:2005}. SWIM model has been implemented in ONE simulator and derivation of SWIM parameters was not clear. Therefore, we try to verify how to derive $\alpha$ value based on node pairs meeting probability in traces and to see how others parameters like neighborhood area should also be adjusted in OMNET++ to get the exact behaviour of mobility patterns using SWIM. The Cambridge traces \cite{Cambridge:2005} consist of different experiments done at different locations. We have used only 3 kinds of traces. They are;

\begin{itemize}
\item INFOCOM 2005: This was an experiment conducted at a conference at Miami inside a hotel in March 2005. This experiment had 41 mobile nodes carried by the attendees of the conference for 4 days.
\item  Cambridge 2005: This was an experiment conducted during October 2005 for 11 days in and around the Cambridge University by distributing iMotes to 36 students and 18 stationary iMotes placed at various locations such as labs, shops, pubs, etc.
\item INFOCOM 2006: This experiment was conducted at a conference (April 2006) at Barcelona in a hotel with 78 attendees as mobile nodes and 20 stationary nodes. The experiment lasted for 5 days.
\end{itemize}

\section{Scenario}
\label{sec:scenario}
Nodes do not attract each other in pure location based models. Locations attract the nodes. This creates a challenge in parameterization of the SWIM model using pairwise contact probabilities. If the nodes attract each other, the pairwise contact probabilities can be used directly to program the attraction between node pairs. A way must be found to translate the location preferences of a pure location based model into the pairwise contact probabilities derived from the real traces. In this section, we discuss an approach to derive the $\alpha$ for the SWIM model from pairwise contact probabilities.

The pairwise contact probabilities are calculated for the whole of the experiment time. The calculation starts with the “number of contacts” matrix which is a NxN matrix, where N is the number of nodes in the experiment. Let us call this matrix as A. The element in the $i^{th}$ row and $j^{th}$ column of the matrix A represents the number of pairwise contacts between the $i^{th}$ and the $j^{th}$ node throughout the experiment. This matrix has the leading diagonal elements as 0 as a node cannot contact itself. The matrix A is symmetric. The algorithm for calculating the pairwise contact probability matrix P from the matrix A is given below in equation (\ref{eq:pl_det}). Pairwise contact probability can be calculated by normalising each element of the matrix A by the sum of the upper triangle of matrix A.

\emph{for each row i in matrix A}

\hskip 2em \emph{for each column j in the row i of matrix A}

\begin{equation}\label{eq:pl_det}
P_{ij}= \frac{A_{ij}}{\frac{\sum_{k=1}^{N} \sum_{l=1}^{N}A_{kl}}{2}}
\end{equation}

\begin{figure}[!ht]
  \centering
    \includegraphics[width=0.45\textwidth]{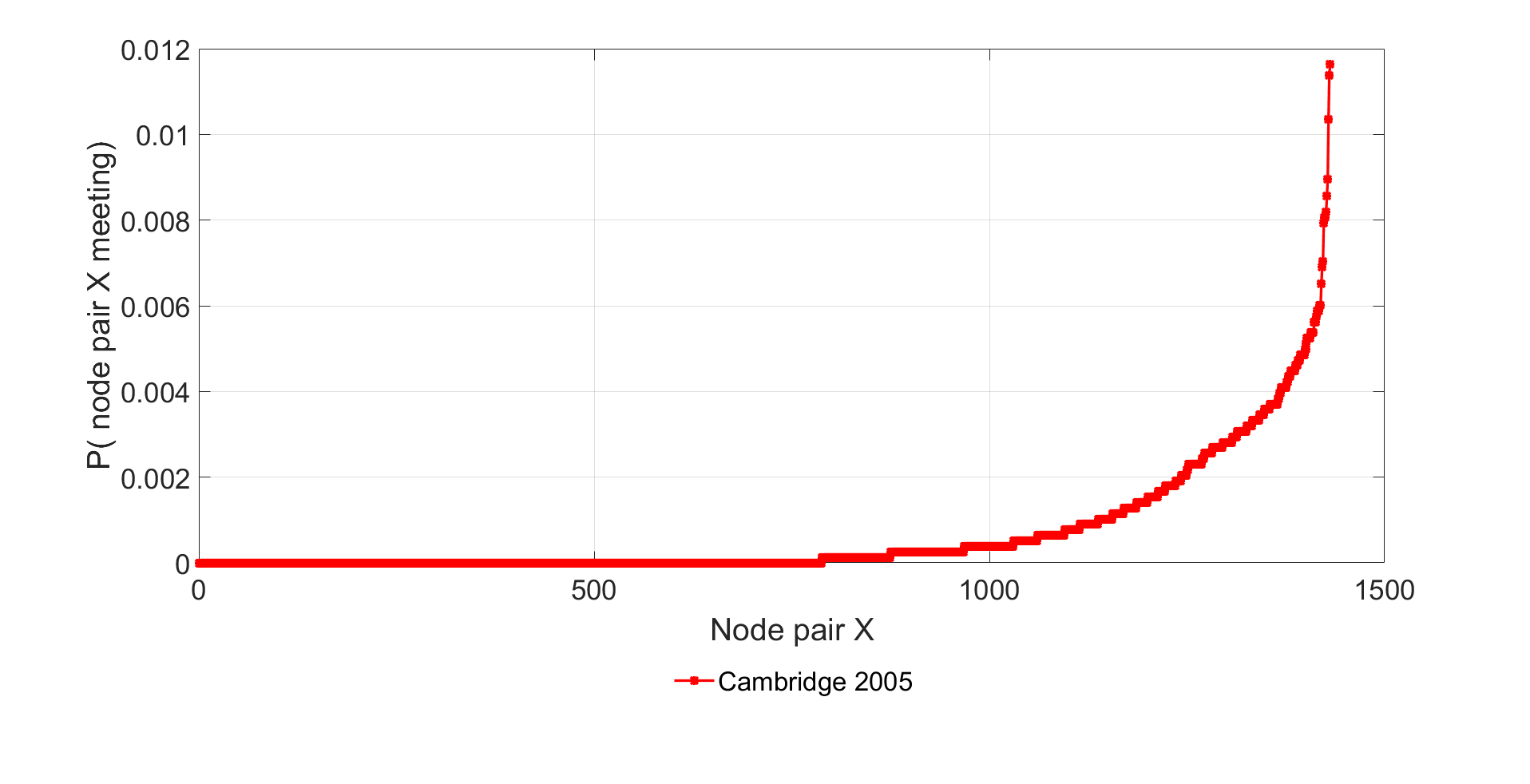}
  \caption{CAMBRIDGE 2005 -  Node pair meeting probability}
  \label{fig:1}
\end{figure}

\begin{figure}[!ht]
  \centering
    \includegraphics[width=0.45\textwidth]{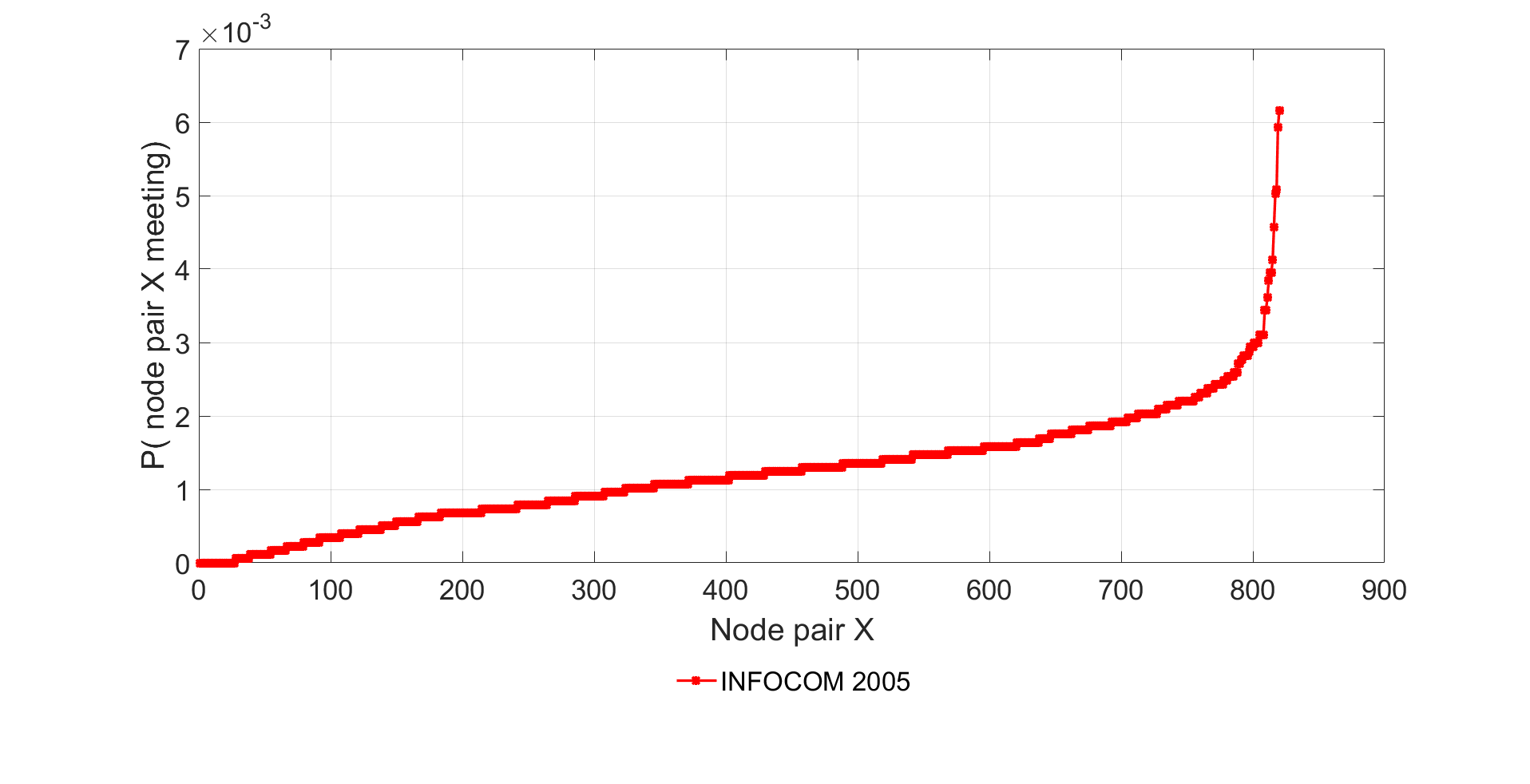}
  \caption{INFOCOM 2005 -  Node pair meeting probability}
  \label{fig:2}
\end{figure}

Deciding the parameter (e.g. $\alpha$ in SWIM model) for a pure location based model from pairwise contact probabilities is not very direct as there is no way to exactly match the values of pairwise contacts between node pairs. Instead, an overall pattern is observed by sorting the upper triangle of the pairwise contact probability matrix P in ascending order. Figure~\ref{fig:1} and figure~\ref{fig:2} show the resultant matrix $P_{pair}$ against the number of possible node pairs obtained from Cambridge and INFOCOM 2005 traces. These figures show two regions, a nearly linear increasing region at the start of the graph which is followed by a sudden increase region which breaks the linearity. 

Higher the probability of visiting only nearby locations, higher will be the probability of preferring only a few nodes. A high $\alpha$ (say 0.9) will make sure that a node visits nearby locations 90 \%  of the time and the other locations outside the neighborhood are visited only 10\% of the time. A low $\alpha$ allows a node to visit outside the neighborhood and hence increasing the probability of meeting all nodes. Based on the above intuitions, the following thumb rule is used to decide on the $\alpha$. 

The more linear the plot increases, lower is the $\alpha$. A fully linear increasing plot nearly parallel to the X axis needs an $\alpha$ of zero to make sure that there is a chance to meet every node. The greater the slope of the non-linear “sudden increase” compared to the linear region, greater the $\alpha$. The above thumb rule is under the assumption that the neighborhood radius with respect to home location of a node is small when compared to the whole map. If neighborhood radius is as big as (or comparable) with the map size, then a high $\alpha$ will not have any meaning since all locations in the map will be treated as nearby locations and the model will start behaving like a random waypoint model for $\alpha$=1. In the Cambridge 2005 trace (figure~\ref{fig:1} ), nearly half the node pairs did not meet each other ($P_{pair}$ = 0) indicating the need for the nodes to stay within the neighborhood and avoid meeting all the nodes. The sudden rise in the $P_{pair}$ (figure~\ref{fig:1}) denotes high probability of meeting for a few node pairs. The Cambridge 2005 trace needs a high $\alpha$ and hence an $\alpha$ of 0.9 is used for simulating the SWIM model.

Greater the $\alpha$, higher the graph rises for higher node pair numbers. For low $\alpha$, the graph will be nearly flat without any non-linear region.

The detailed comparision of results are shown here with regard to only the Cambridge traces. The proposed $\alpha$ of 0.9 is the nearest match for the Cambridge 2005 trace. An $\alpha$ of 1 might also be the best match as it would reach even higher than $\alpha$ = 0.9. But the task is just to find out the way in which the pairwise node meeting probability behaves for different $\alpha$ values. 

Table~\ref{tab:parameter_values} lists the parameters used to configure a scenario with SWIM mobility model based on Cambridge trace.

\begin{table}[!ht]
	\centering
	\footnotesize
	\begin{tabular}{|p{4cm}|p{4cm}|}
		\hline
		\multicolumn{1}{|c|}{\textbf{Parameter}} & \multicolumn{1}{|c|}{\textbf{Cambridge Traces}} \\
		\hline
		\hline
		Number of nodes        & 36 mobile nodes, 18 stationary nodes \\
		\hline
		initialX, initialY     & Randomly selected values \\
		\hline
		maxAreaX, maxAreaY     & 2000 meters x 2000 meters \\
		\hline
		waitTime               & exponential (30 minutes) \\
		\hline
		alpha ($\alpha$)                 & 0.9 \\
		\hline
		noOfLocations          & 38 (Stationry nodes are placed in the locations randomly.) \\
		\hline
		Beacon Interval 	& Mobile nodes (10 min), 4 long range nodes (2 min), 2 short range nodes (6 min), 12 short range nodes (10 min)    \\
		\hline
		Radio range         & Mobile nodes (11m), Stationary - short (11m), long (22m) \\
		\hline
		Simulation time      & 11 days \\
		\hline
		Neighborhood radius      & 100m \\
		\hline
	\end{tabular}
	\vspace{0.1cm}
	\caption{Used Parameter Values}
	\label{tab:parameter_values}
	\vspace{-0.6cm}
\end{table}

In this work, we analyse the “node pair meeting probability” which gives us a measure of how often node pairs meet. The node pairs do not indicate any particular node pair. Instead, the node pair numbers in the X axis just indicate the number of node pairs in the simulation as shown in figure 1, 2 and 3. The node pair meeting probability used in this work is useful in calculating the data dissemination time similar as in \cite{KAMINI:2015}. Assuming that we randomly assign exact node pair identities to the X-axis, the data dissemination time will not only depend upon contact duration and inter contact time, but, it also depends on the node which is met next. This is due to the simple fact that each node might have a different set of data at any given point of time and due to this, the node which will be met next will play a major role in data dissemination time. For example, in a situation where a Node-X meet only one other Node-Y during the whole experiment, it is totally possible that the Node-X will never have all the data in the network and the network may never converge towards a finite data dissemination time due to this Node-X. 

It is important to note that, since we only try to analyse the effect of $\alpha$ values on pairwise contact probabilities, the other parameters are fixed such as waiting time parameters. The parameters other than $\alpha$ which are set as constant cannot represent the human mobility in the traces very accurately because they are derived from the aggregate data of all nodes in the mobility traces. Therefore, in this work, we only try to tune the most important deciding parameter of the SWIM model ($\alpha$) using the pairwise contact probability.

\section{Analysis of Results}
\label{sec:swim_eval} 

\begin{figure}[!ht]
  \centering
    \includegraphics[width=0.45\textwidth]{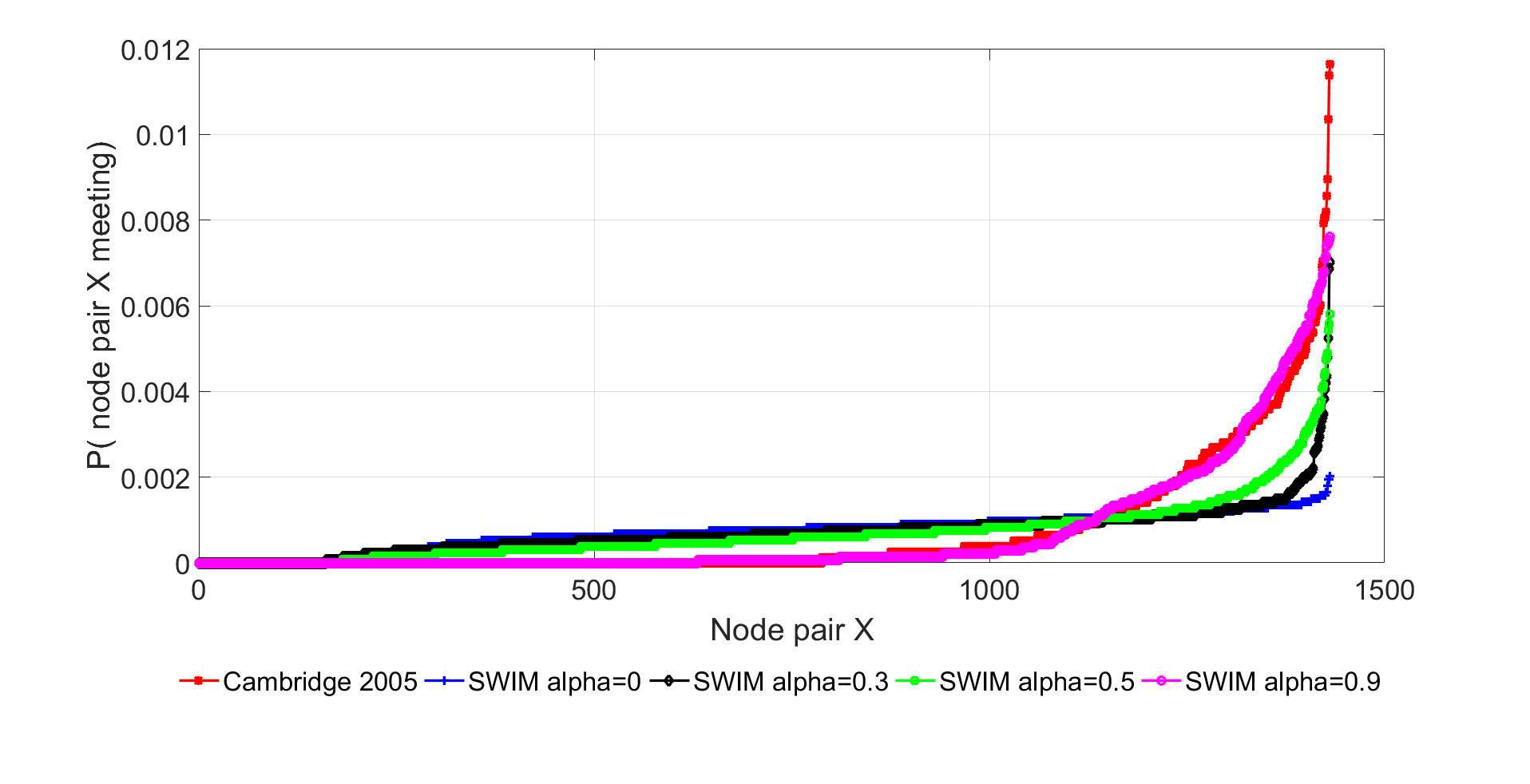}
  \caption{Node pair meeting probability - Cambridge vs SWIM}
  \label{fig:29}
\end{figure}

\begin{figure}[!ht]
  \centering
    \includegraphics[width=0.45\textwidth]{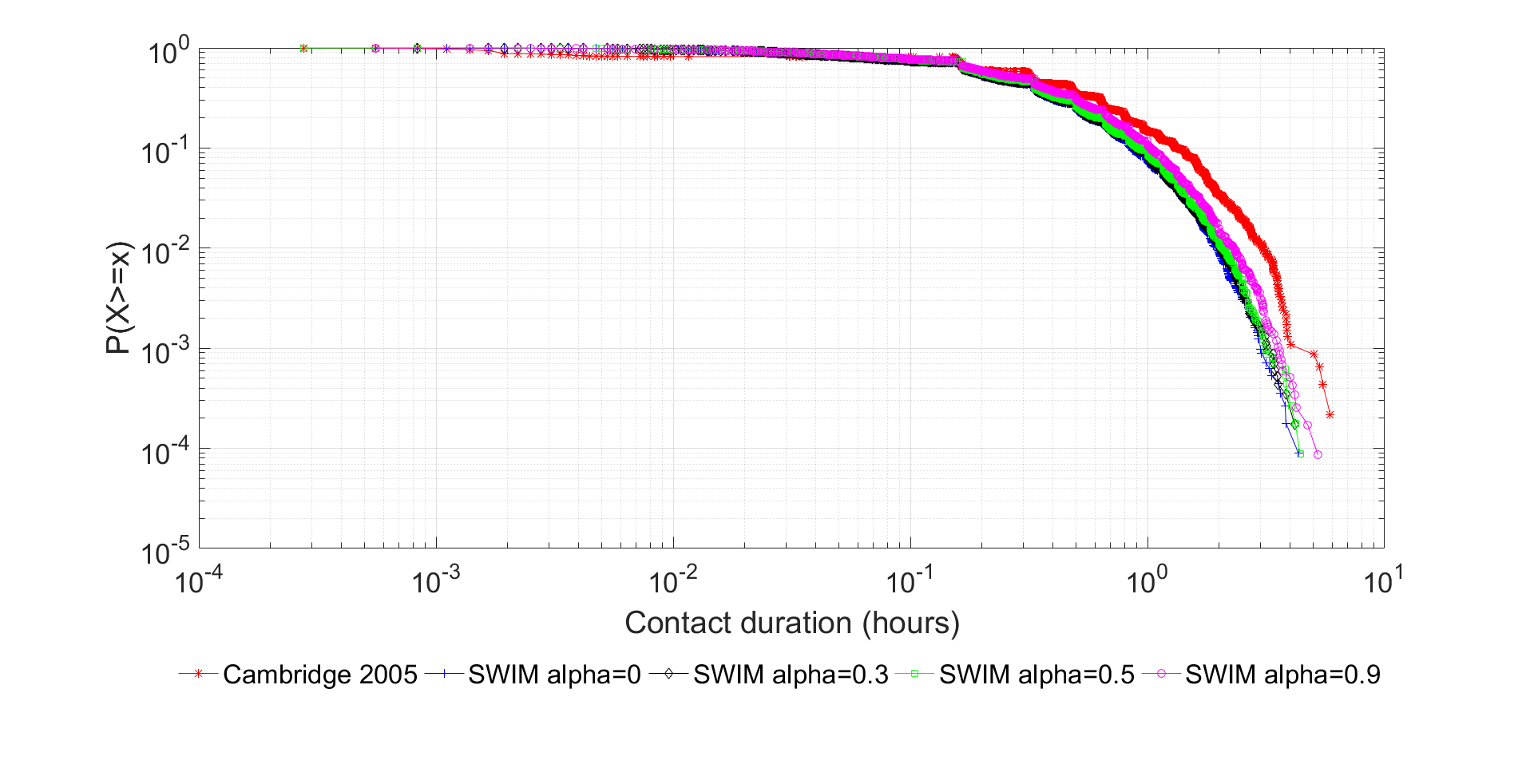}
  \caption{Contact duration - Cambridge vs SWIM (log-log axis)}
  \label{fig:30}
\end{figure}

The Cambridge 2005 trace is simulated with multiple $\alpha$ values and the parameters mentioned in Table~\ref{tab:parameter_values}. Figure~\ref{fig:29} shows that the proposed $\alpha$ of 0.9 matches the node pair meeting probability of Cambridge 2005 trace. The greater the $\alpha$, higher the graph rises for higher node pair numbers. For low $\alpha$ values, the graph is nearly flat without any non-linear region. Figure~\ref{fig:30} shows that the contact durations of Cambridge 2005 and SWIM model are almost a perfect match. Two nodes will be in contact if they are at rest at the same place within communication range or if they meet while in motion.

\begin{figure}[!ht]
  \centering
    \includegraphics[width=0.45\textwidth]{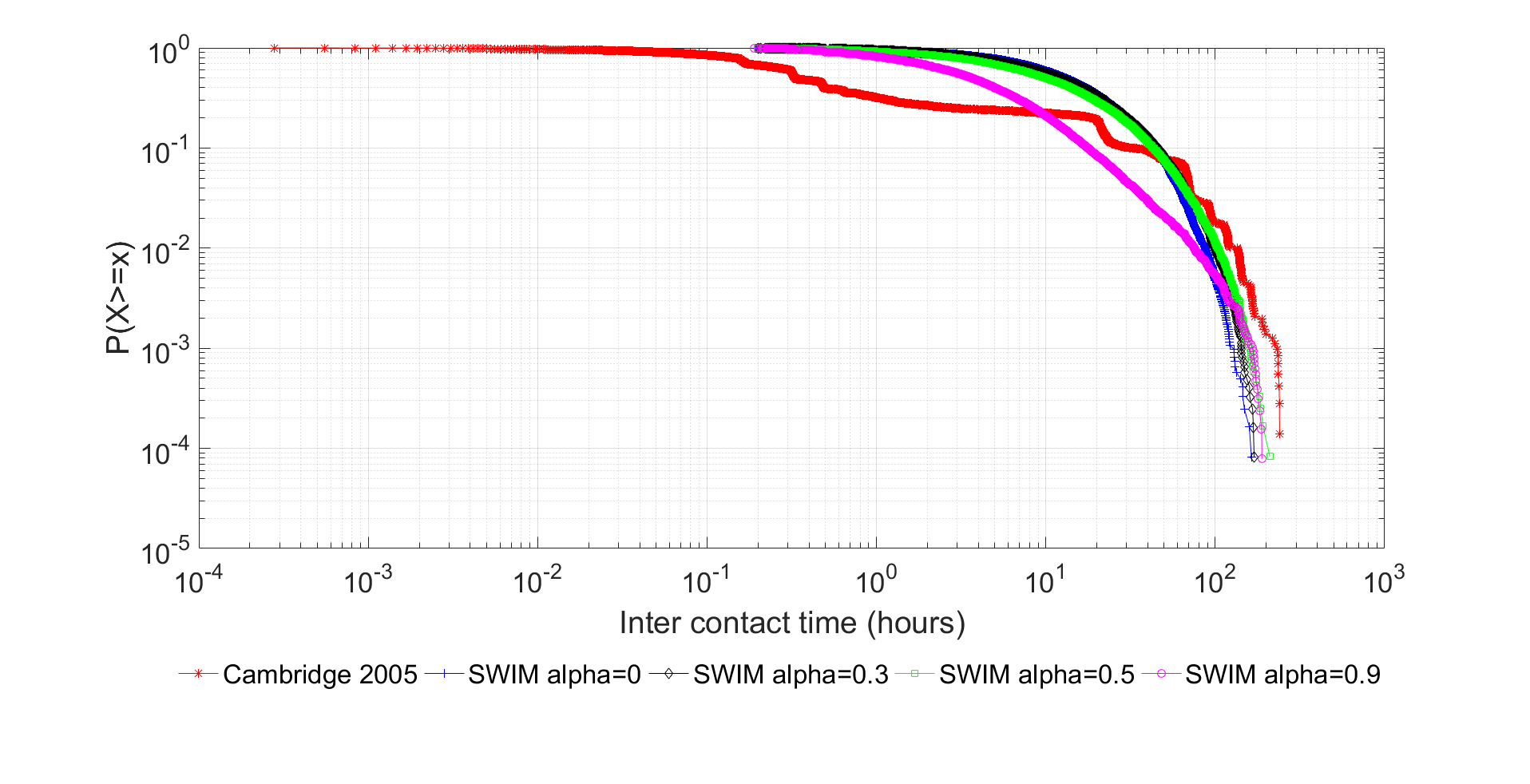}
  \caption{Inter-contact times - Cambridge vs SWIM (log-log axis)}
  \label{fig:31}
\end{figure}

The latter is not easily controllable as it depends on a lot of random movements. The wait times are the easiest way to tune the contact durations. Choosing the right distribution for the wait times plays a key role in achieving a good match for the contact durations. In SWIM model, since the nodes move with speed equal to the distance between the source and destination, the motion of the nodes contribute negligibly to the contact durations. The wait times are the main contributors to contact durations. 

\begin{figure}[!ht]
  \centering
    \includegraphics[width=0.45\textwidth]{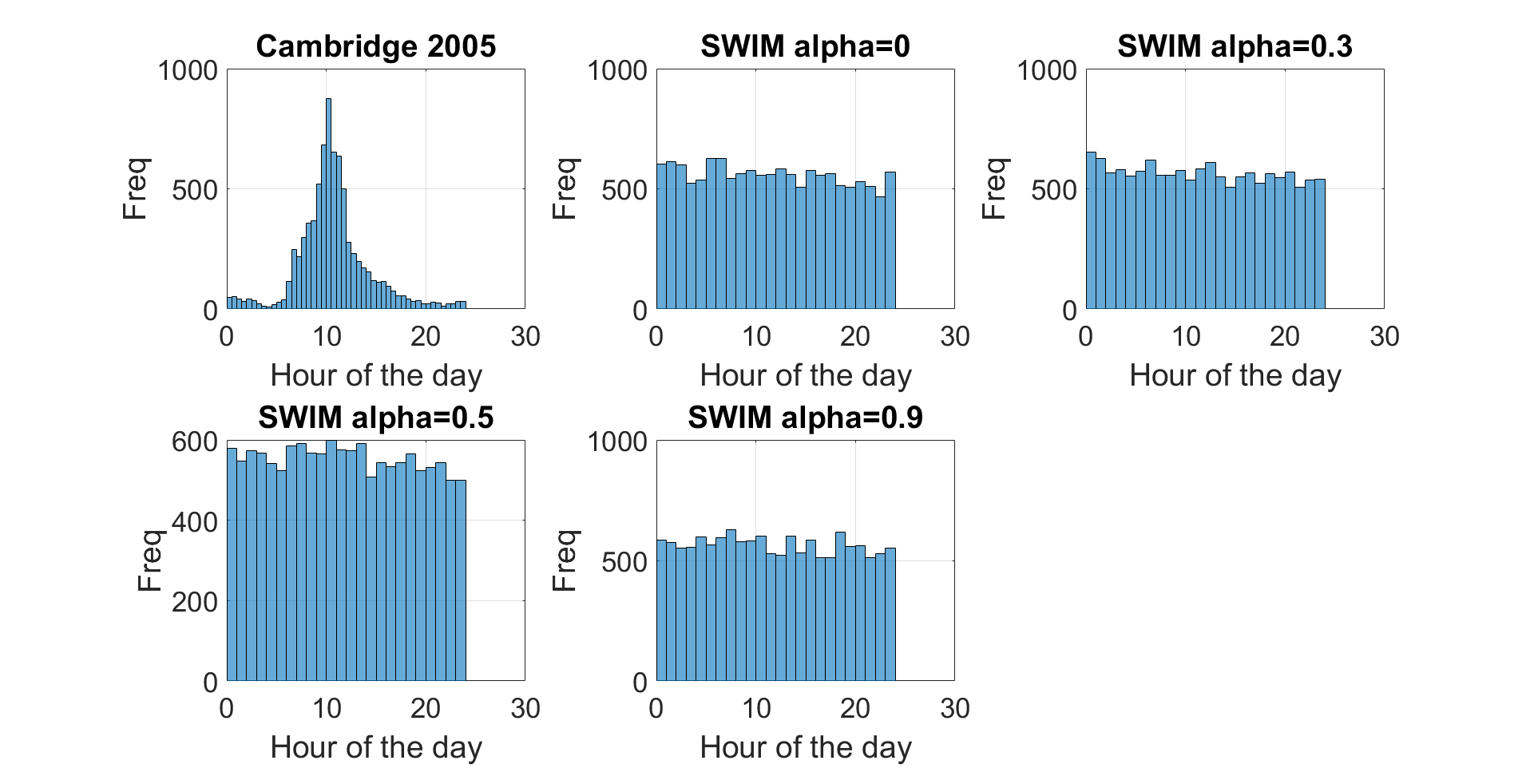}
  \caption{Number of overall pairwise contacts based on hour of the day}
  \label{fig:32}
\end{figure}

In figure~\ref{fig:31}, the inter-contact times of SWIM simulation are very similar to the Cambridge 2005 trace. The graph follows a power law up to approximately 12 hours and followed by an exponential cut-off. In figure~\ref{fig:32}, we can see the day and night patterns of Cambridge 2005 and the absence of it in SWIM model. 

\begin{figure}[!ht]
  \centering
    \includegraphics[width=0.45\textwidth]{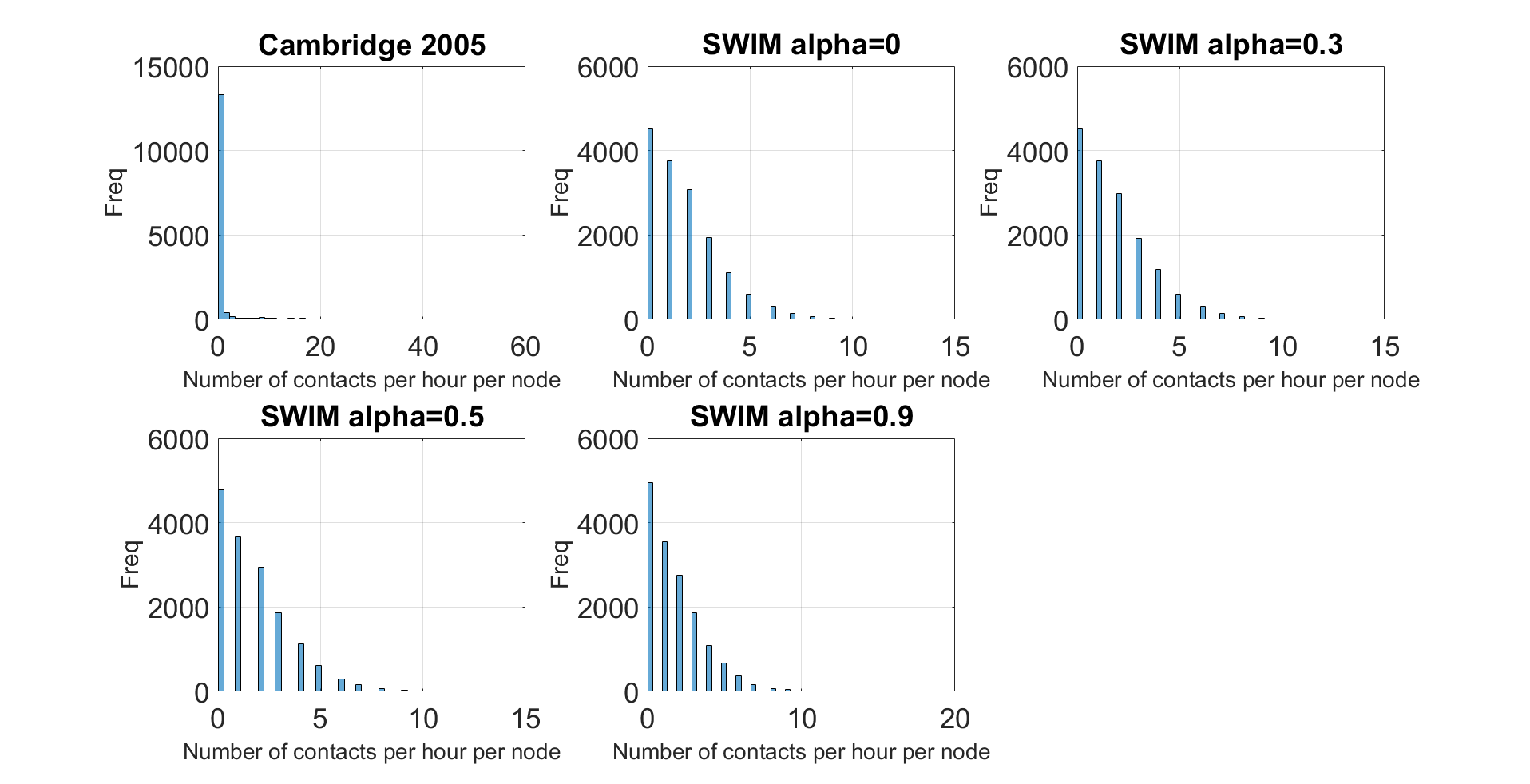}
  \caption{Aggregate number of contacts per hour per node}
  \label{fig:33}
\end{figure}

Figure~\ref{fig:33} shows the frequency of a node having zero number of contacts per hour is very high for Cambridge 2005 when compared to the SWIM simulations. To have zero contacts per hour per node, a node must not meet any node for one hour. This is highly probable only if a node is at rest. Movement increases the possibility of meeting another node. Since there is no assurance of rest in a periodic manner, the possibility of SWIM model having zero contacts per hour per node is not as high as Cambridge 2005. There is a lack of high number of contacts per hour per node in SWIM. In Cambridge 2005 trace in figure~\ref{fig:1}, some of the nodes have achieved more than 20 contacts per hour even though the frequency is very low. But, SWIM simulations did not achieve more than 10 contacts per hour per node for any $\alpha$ value. In real life traces, nodes have different wait time parameters and different $\alpha$ values. If the simulation is done with just one common $\alpha$ and wait time distribution for all the nodes, then it will not be possible to achieve low and high contacts per hour per node. To achieve zero contacts per hour per node, there must be a possibility to have longer wait times which enable the nodes to be less active. The lack of a night time removes the possibility of the nodes resting in a periodic manner which in turn reduces the possibility of having longer wait times. To achieve more contacts per hour, a node must either move quickly without waiting long at any location to meet other nodes within short durations or a node must be visited by a lots of other nodes within a short amount of time. This is possible only if the nodes are programmed in a heterogeneous manner in terms of wait time and $\alpha$.

\section{Conclusion}
\label{sec:conclusion}

In this work, pairwise contact probabilities are used to guess the alpha value. However, we do not yet have a mathematical method to derive an alpha value using the pairwise contact probabilities, we try to analyse the pattern in node pair meeting probability for different alpha values and propose a thumb rule for guessing an alpha value. The work in \cite{Mei:2009} introduces the SWIM model but the reason for alpha value to be chosen is not very clear. The unique contribution of this work comparing to the work in \cite{Mei:2009} is: 
\begin{itemize}
\item We find a way to parameterize a location based model (SWIM) using pairwise contact probabilities extracted from real traces. In a location based model, we do not control the attraction between node. On the other hand, we have control over where a node will go using location preferences. In this work, we try to translate location preferences (neighborhood area and $\alpha$) into node preference by sorting the pairwise contact probabilities. In this sorting process, node pair identities are lost and we use the remaining pattern to parameterize the SWIM model. 
\end{itemize}

This work investigated how to tune SWIM parameters to replicate the pairwise contact probabilities of a very heterogeneous real life traces of Cambridge traces. The $\alpha$ values and the size of the neighborhood area have an effect on the mobility of nodes.  A larger $\alpha$ value results in selecting target locations close to the home location which in turn limits the number of nodes encountered. On the other hand, a smaller $\alpha$ value results in selecting popular locations which in turn increases the probability to meet all the nodes in the experiment.  

Pairwise contact probabilities are parameters of both social and location attractions. Some people move to meet other people and some people move to go to places. The use a purely location based model like SWIM to match the pairwise contact probabilities of real life movements is a hard challenge. By design, SWIM model still lacks heterogeneity in terms of activity level, waiting times and other parameters such as neighbourhood radius and  popularity decision threshold. Therefore, some of the further work include:

\begin{itemize}
	\item Different $\alpha$ values:  can be used to represent day and night time mobility behaviour. Nodes can be divided into clusters representing a particular behaviour with regard to $\alpha$ (e.g., behaviour of students vs teachers)
	\item Different neighbourhood radius: multiple sectors with different radii making each sector having a different priority of visiting. This allows the possibility of expanding the neighbourhood into fine divisions (e.g. Kitchen and lab area in Cambridge traces)
	\item  A mathematical model: This work is starting point for using pairwise contact probabilities for parameterizing the SWIM model. Therefore, a mathematical model to predict the $\alpha$ and other parameters of SWIM model based on existing properties of real life traces (e.g. pairwise contact probability) is a part of future work. The approach presented in this paper is purely graphical in guessing the $\alpha$ for simulating the SWIM model. 
\end{itemize}

\bibliographystyle{IEEEtran}
%

\end{document}